\documentclass[
 ,final            		
  ,numberedheadings 	
  ]
 {aipproc}
\usepackage{amsmath,amsthm}
\layoutstyle{8d}

\begin{document}

\title{On the \emph{Google}-Fame of Scientists and Other Populations}

\author{James P.~Bagrow}{
  address={Department of Physics, Clarkson University, Potsdam NY 13699-5820}
}
\author{Daniel ben-Avraham}{
   address={Department of Physics, Clarkson University, Potsdam NY 13699-5820}
}

\begin{abstract}
We study the fame distribution of scientists and other social groups as measured by the number of
\emph{Google} hits garnered by individuals in the population.  Past studies have found that the fame distribution decays either in power-law~\cite{originalAces} or exponential~\cite{ourPrevious} fashion, depending on whether individuals in the social group in question enjoy true fame or not.  In our present study we examine critically \emph{Google} counts as well as the methods of data analysis.  
While the previous findings are corroborated in our present study, we find that, in most situations, the data available does not allow for sharp conclusions.
\end{abstract}

\maketitle


\section{Introduction}

The concept of Fame within a population has critical social and economic impact.  Recently, the idea of using the number of \emph{hits} returned from a search of a person's name on \emph{Google} as a means of quantifying that person's fame has been explored \cite{originalAces,ourPrevious}.  A seminal paper explored the fame of a unique population, that of World War I ``ace'' pilots \cite{originalAces}, and found, among other things, a power-law decay in the tail of the distribution.  More recent work~\cite{ourPrevious}  has applied this to a population of scientists who have published on the \emph{cond-mat} e-print archive\footnote{\url{http://arxiv.org/archive/cond-mat}}.  The tail of their fame distribution was best fit by an exponential.  On the other hand, the fame of other populations was found to follow a power-law decay.  The difference was attributed to the fact that scientists habitually use the World Wide Web as a professional means of communication and cite each other on the web in relation to their published work.

\emph{Google}'s goal as a service is to provide accurate search results to its users.   For the purposes of determining a subject's fame, what is most relevant is not having accurate results listed first, as it is for most users, but to have an accurate \emph{count} of those results.  Unfortunately, \emph{Google} does not provide enough accuracy, and there are several reasons for this \cite{googleProblems}.   

\emph{Google} acknowledges that the hits count given is an estimate, but does not elaborate on the accuracy of this estimation nor reveal how it is calculated.  It seems reasonable to assume that very small counts are more accurate than larger ones.  This means that the error is largest in the tail of the fame distribution, and it is this region that is of most interest.  In addition, the tail of the distribution is more likely to contain results that are over-counted, further compounding the error.  

In~\cite{originalAces}, over-counting was prevented by verifying each hit by hand, a time-consuming procedure that limited the sample size.  At the time of this writing, \emph{Google} only returns the first 1000 hits, so it is impossible to verify the accuracy of any results beyond that number, and one must trust in \emph{Google's} estimation.  Even manual verification is limited.

The previous searches in \cite{originalAces,ourPrevious} used a search lexicon including the boolean \texttt{OR} operator.  We have since found out that Google returns incorrect hit counts when \texttt{OR} is used \cite{googleProblems}.  For a simple illustration, a search for \texttt{cars OR automobiles} returns 80.5 million hits (at the time of this writing) while searches for \texttt{cars} and \texttt{automobiles} return 94.2 million and 8.82 million hits, respectively, violating basic set theory.  Thus, the previous work must be reproduced using a better lexicon.  In the current work, all our searches avoid the problematic \texttt{OR} operator.  See Table \ref{table:results}.

Despite these issues, \emph{Google} still provides an excellent tool for research.  It is the simplest means of getting the most information available and it commands a very large sample space.  For example, the work in \cite{googleSemantic} uses hit counts to ``teach'' the semantic meaning of words to software --- a central problem in Artificial Intelligence.  Related words such as `painter' and `artist' will have many more joint occurrences than disparate words, such as `plumber' and `artist', leading to higher hit counts.  Their work confirms that \emph{Google} yields reasonable results when avoiding the \texttt{OR} and using the \texttt{AND} operator only.  

\emph{Google} has been generous enough to open their search interface to allow tools to be created that can perform \emph{Google} searches automatically\footnote{\url{http://code.google.com}}.  We have used this to eliminate the laborious task of entering single queries and recording the hits count.  Larger populations can be searched much more quickly using an automatic tool.  For the present work, we used a script that performed the searches from a web server.  An easier way still is with the open-source \emph{PyGoogle} package\footnote{\url{http://pygoogle.sourceforge.net}}, which integrates the \emph{Google} search interface with the Python programming language.  

In both \cite{originalAces} and \cite{ourPrevious}, fits were performed by binning the search results with exponentially-sized bins and then fitting to the binned data using least-squares.  A better technique than binning, when working with sparse data, is examining cumulative distributions~\cite{newmanPower}, as we do in the present work.  Also, it has been shown that there are problems with using  least-squares fits to logarithmic plots~\cite{powerProblems}.  One problem is that the log operation magnifies the error in the tail. Least-squares fitting assumes that errors for each data point are uniform and will not properly weigh the noisier tail.   In this work, we use a more robust technique, that of Maximum Likelihood, to achieve less biased fits.  See section \ref{section:maxLike}.  

All distributions studied here exhibit a power-law tail, although for many the tail covers a very narrow range.  For the scientists populations, we observe  a power-law tail only in the top 12\% of the data.  Most of the data for scientists is best fit by an exponential, just as found in the previous study~\cite{ourPrevious}.   In contrast, other populations do not fit to an exponential over any sizable range.  See section \ref{results}.

\section{Maximum Likelihood Fitting} \label{section:maxLike}

A better technique to determine the parameter(s) of a probability distribution from sampled data is that of Maximum Likelihood. The results are more robust in terms of error-weighing.  This is a very common technique and is covered in many statistics and regression texts.

To briefly illustrate how maximum likelihood works, let us derive the \emph{Maximum Likelihood Estimator} (MLE) for $\lambda$, the parameter for an exponential probability distribution:
\begin{equation}
P(x) = \lambda e^{-\lambda x}
\end{equation}
The goal of Maximum Likelihood is to find the \emph{most likely} $\lambda$ given the existing data.  For this, we start with the probability of the experiment given $\lambda$, assuming independence of the data points:
\begin{align}
P\left( x_1, x_2, ... x_N | \lambda \right)  & =  \prod_{i=1}^{N} \lambda e^{-\lambda x_i} \\
& =  \lambda^N \exp \left(-\lambda \sum_{i=1}^{N} x_i \right) 
\end{align}
where $x_i$ is the unbinned data gathered from the experiment.  This function is the total probability of all measurements occurring in the experiment.  From this, we define the \emph{likelihood function}, using Bayes' Theorem:
\begin{equation}
l(\lambda | x_1, x_2, ...x_N ) = P(x_1, x_2, ...x_N | \lambda ) \frac{P(\lambda)}{P(x_1, x_2, ...x_N )}
\end{equation}
This is the \emph{likelihood} of $\lambda$ given the experimental data.  Assuming $P(x_1, x_2, ...x_N )=1$ (the experiment has already occurred) and $P(\lambda)$ is uniformly distributed (all $\lambda$'s are equally likely), then $l(\lambda | x ) \propto P(x | \lambda )$.  To find the most likely $\lambda$, we must find the maximum of this function with respect to the parameter $\lambda$.  To simplify the calculation, we will instead maximize the log-likelihood function, $L$, which is equivalent:
\begin{align}
L = \ln (l ) & = N \ln \lambda - \lambda \sum_{i=1}^{N} x_i \\
\frac{dL}{d \lambda} & = \frac{N}{\lambda} - \sum_{i=1}^{N} x_i  = 0 \\
\lambda & = N / \left( \sum_{i=1}^{N} x_i \right) \label{eqn:mle_exp}
\end{align}
This is just the inverse of the mean, exactly as expected for an exponential distribution.   We need not account for the proportionality between $l(\lambda | x )$ and $P(x | \lambda)$ because we only used the derivative of $L$ 

The other probability distribution we are concerned with is the power-law distribution:
\begin{equation}
P(x) \propto x^{-\gamma}
\label{powerLaw}
\end{equation}
For the MLE of $\gamma$, we reproduce the derivation given in \cite{newmanPower}. The first step is to normalize Eqn.~(\ref{powerLaw}) for the given data points:
\begin{equation}
P(x) = C x^{-\gamma} = \frac{\gamma-1}{x_{\text{min}}} \left( \frac{x}{x_{\text{min}}} \right)^{-\gamma}
\end{equation}
where $C$ is a constant of proportionality and $x_{\text{min}}$ is the smallest data point from the given sample.   This is then used to get the probability of the experiment:
\begin{equation}
P(x_1, x_2, ...x_N | \gamma ) = \prod_{i=1}^N P(x_i) = \prod_{i=1}^N \frac{\gamma-1}{x_{\text{min}}} \left( \frac{x_i}{x_{\text{min}}} \right)^{-\gamma}
\end{equation}
This is proportional to $l(\gamma | x)$ as before:
\begin{equation}
l(\gamma | x_1, x_2, ...x_N ) = \prod_{i=1}^N \frac{\gamma-1}{x_{\text{min}}} \left( \frac{x_i}{x_{\text{min}}} \right)^{-\gamma}
\end{equation}
Again, we work with $L = \ln l$, which is equivalent for finding the most likely $\gamma$.  Then:
\begin{align}
L & = \sum_{i=1}^N \left(   \ln(\gamma-1) - \ln x_{\text{min}} - \gamma \ln \frac{x_i}{x_{\text{min}}}       \right) \\
& = N \ln ( \gamma -1 ) - N\ln x_{\text{min}} - \gamma \sum_{i=1}^N \ln \frac{x_i}{x_{\text{min}}} 
\end{align}
The MLE for $\gamma$ can then be found:
\begin{align}
\frac{dL}{d\gamma} & = \frac{N}{\gamma -1} - \sum_{i=1}^N \ln \frac{x_i}{x_{\text{min}}} = 0 \\
\gamma & = 1 + N / \left( \sum_{i=1}^N \ln \frac{x_i}{x_{\text{min}}}\right) \label{eqn:mle_pow}
\end{align} 

 Maximum Likelihood derives an estimator for a distribution's parameter(s),    regardless of whether the sampled data truly does come from such a distribution.  Hence, one needs a way to test how well the estimator matches the sample.  For our purposes, the Kolmogorov-Smirnov (KS) Test works quite well \cite{powerProblems}.   This test compares the cumulative distribution function (CDF) of the hypothesized probability distribution to the empirical CDF of the sampled data.  The test statistic is:
 \begin{equation}
 K = \sup_x \left| F(x) - S(x) \right|,
 \end{equation}
 where $F(x)$ is the hypothesized CDF and S(x) is the empirical CDF.  $K$ is then compared with a critical value (for the given significance level) which can be found in a table or generated by software.  MATLAB's Statistics Toolbox has a built-in KS-Test function, \texttt{kstest()}.

\section{Populations}
We have been able to greatly expand upon the number of searches performed compared to previous work.  In addition, due to the problems with the \texttt{OR} operator, we have performed multiple searches of the same population using progressively inclusive lexicons.  Here we describe the populations studied.

\begin{description}
\item[Scientists:]
 Two populations of scientists were used in this study.  The smaller one (of size 449) is the same population used in \cite{ourPrevious}.  The larger population (of size 1625) is a list of authors who have published recently on cond-mat and was harvested using arXiv's OAI XML feed\footnote{See \url{www.openarchives.org}}.
 
 \item[Aces:]
 The population of 1851 aces contains the 393 German aces studied in \cite{originalAces} as well as all the listed aces of other nationalities\footnote{See \url{www.theaerodrome.com}}.  
 
\item[Actors:]
 The actors population contains 778 actors who were born on the second or third of each month between the years 1950 and 1955, as collected from the archives of the Internet Movie Database\footnote{\url{www.imdb.com}}.  These selection criteria were chosen to insure a mostly uniform sample and to give all the chosen subjects roughly the same career length.
 
 \item[Villains:]
 The villains population was gathered from a user-contributed list of antagonists from fictional media\footnote{See \url{en.wikipedia.org/wiki/List_of_villains}}.  This list contains both fictitious characters and real people who have appeared in fictional works.  Since this list was generated by users, the  characters must already enjoy a substantial level of popularity.  
 
\item[Programmers:]
 Similar to the villains population, this population was collected from a user-contributed list of famous programmers\footnote{See \url{en.wikipedia.org/wiki/List_of_programmers}}; people who have made a large contribution to computing, the Internet, etc., such as Tim Berners-Lee, who invented the World Wide Web, and Bill Gates, a co-founder of Microsoft. As with the villains, it seems safe to assume that this population is ``famous''. 
 
\item[Clarkson Students:]
 The students population was chosen from Clarkson University's student directory.  It consists of all students (undergraduate and graduate) whose last name contains the letter ``e''.  This criterion was chosen simply to make it easy to harvest a large collection of names from the online student directory.  We assume this is a ``non-famous'' population, in that the students are too young to have amassed any real fame.  
 
\item[Runners:]
 This population was used previously \cite{ourPrevious}.  The original searches used the erroneous \texttt{OR} operator and are here reproduced without it.
 
\end{description}

\section{Results and Analysis} \label{results}

Table \ref{table:results} contains the power-law exponents and search lexicons for the populations studied.  Many of the power-law exponents are $\approx2$, as first predicted in \cite{originalAces}.   All populations display a power-law tail, regardless of whether they are ``famous'' or not.  It should be pointed out, however, that for some populations the range fitting a power-law is extremely narrow, casting doubt on this interpretation.  In those cases, an exponential distribution may fit as well.  Most of the scientists distributions fit an exponential over much of the ``non-tail''.  See Table \ref{table:resultsSCI}.  Clarkson students, another population assumed to be non-famous, does not fit to an exponential over such a range.  This is further evidence that the exponential distribution for scientists stems from their use of the World Wide Web as a professional means for disseminating research, rather than related to fame.

\begin{table}
	\begin{tabular}{rcccl}
	\hline
	\tablehead{1}{c}{b}{Population (Size)}
	  & \tablehead{1}{c}{b}{Search}
	  & \tablehead{1}{c}{b}{$\gamma$}
	  & \tablehead{1}{c}{b}{Fitting Range\tablenote{For example, Top 99 means that the fit was applied to only the 99 highest-ranked searches.  Note that some sub-samples constitute less than 10 percent of the population and that the tail contains the noisiest, and therefore least reliable, data.}}  
	  & \tablehead{1}{c}{b}{Lexicon}    \\
	\hline
	 Scientists (449) 	& 1 & 1.82   	& Top 99 		&  <name>     \\
					& 2 &  2.18 	&  " "			& <name> physics \\
					& 3 & 2.29 	&   " " 		& <name>  statistical  physics \\
	 				& 4 & 2.69 	&   "  " 		& <name> statistical physics condensed \\
	 \hline
	 Scientists (1625)  	& 1 & 2.02  	&  Top 105 	&  <name>   \\
	 				& 2 & 1.77  	& Top 240 	& <name>  physics  \\
					& 3 & 2.08  	& Top 150		& <name>  statistical  physics \\
					& 4 & 2.00  	&   Top 210 	& <name> statistical physics condensed \\
	\hline
	Aces (1851)		& 1 &  2.74   	&  Top 99 		&  <name> WWI  \\
					& 2 & 3.62 	& " "  			&  <name> WWI ace \\
	\hline
	Actors (778) 		& 1 & 1.88  	& Top 120   	& <name> \\
					& 2 & 2.04 	& " " 			&  <name> movie \\
					& 3 & 2.10   	& " "  			&  <name> movie actor \\
	\hline
	Villains (421) 		& 1 & 1.57  	&  Top 99 		&  <name> \\
					& 2 & 1.86   	& " "			& <name> villain \\
					& 3 & 2.03   	&  " " 			&  <name> villain evil\\
	\hline
	Programmers 		& 1 & 1.88  	& Top 59 		& <name>  \\
	(148) 			& 2 & 2.16 	&  " " 			& <name> programmer \\
					& 3 & 2.03   	&  " " 			& <name> computer \\
					& 4 & 2.43   	&  " "  		& <name> computer programmer\\
	\hline
	Clarkson Students  	& 1 & 1.74  	&  Top 119  	& <name>  \\
	(1533) 			& 2 & 1.99  	&  " " 			& <name> clarkson\\
					& 3 & 2.57  	& " "  			& <name> "clarkson university"\\
	\hline
	Runners (222) 		& 1 & 1.71   	&  Top 99  	& <name>  \\
					& 2 & 1.92    	& " " 			&  <name> olympics \\
	\hline
	\end{tabular}
	\caption{MLE Power-Law Fits to Search Results.  All fits pass the KS-test ($\alpha=0.05$).}
	\label{table:results}
\end{table}

\begin{table}
	\begin{tabular}{rcrccc}
	\hline
	\tablehead{1}{c}{b}{Population (Size)}
	  & \tablehead{1}{c}{b}{Search}
	  & \tablehead{1}{c}{b}{$\lambda^{-1}$}
	  & \tablehead{1}{c}{b}{$K$}  
	    & \tablehead{1}{c}{b}{$CV$\tablenote{ The critical value that is compared to $K$.  A distribution passes the KS-test when $K < CV$.}}
	  & \tablehead{1}{c}{b}{Fitting Range}    \\
	\hline
	 Scientists (449) 	& 1 & 1040   	& 0.0877 		& 0.0908	&  230 - 449    \\
					& 2 &  591 	&  0.0663		& 0.0706	&  85 - 449 \\
					& 3 & 385  	&  0.0650		& 0.0688	&  65 - 449 \\
	 				& 4 & 134  	&  0.0602 		& 0.0644	&  10 - 449 \\
	 \hline
	 Scientists (1625)  	& 1 & 468  	& 0.0495 		& 0.0505	& 910 - 1625     \\
	 				& 2 & 343  	& 0.0385		& 0.0395	& 455 - 1625  \\
					& 3 & 161  	& 0.0397		& 0.0416	& 570 - 1625  \\
					& 4 & 75		& 0.0423  		& 0.0369	& 280 - 1625 \\
	\hline

	\end{tabular}
	\caption{MLE Exponential Fits to Search Results.  All fits pass the KS-test ($\alpha=0.05$) except for Scientists (1625) Search 4.}
	\label{table:resultsSCI}
\end{table}

\begin{figure}
  \includegraphics[width=1.15\textwidth]{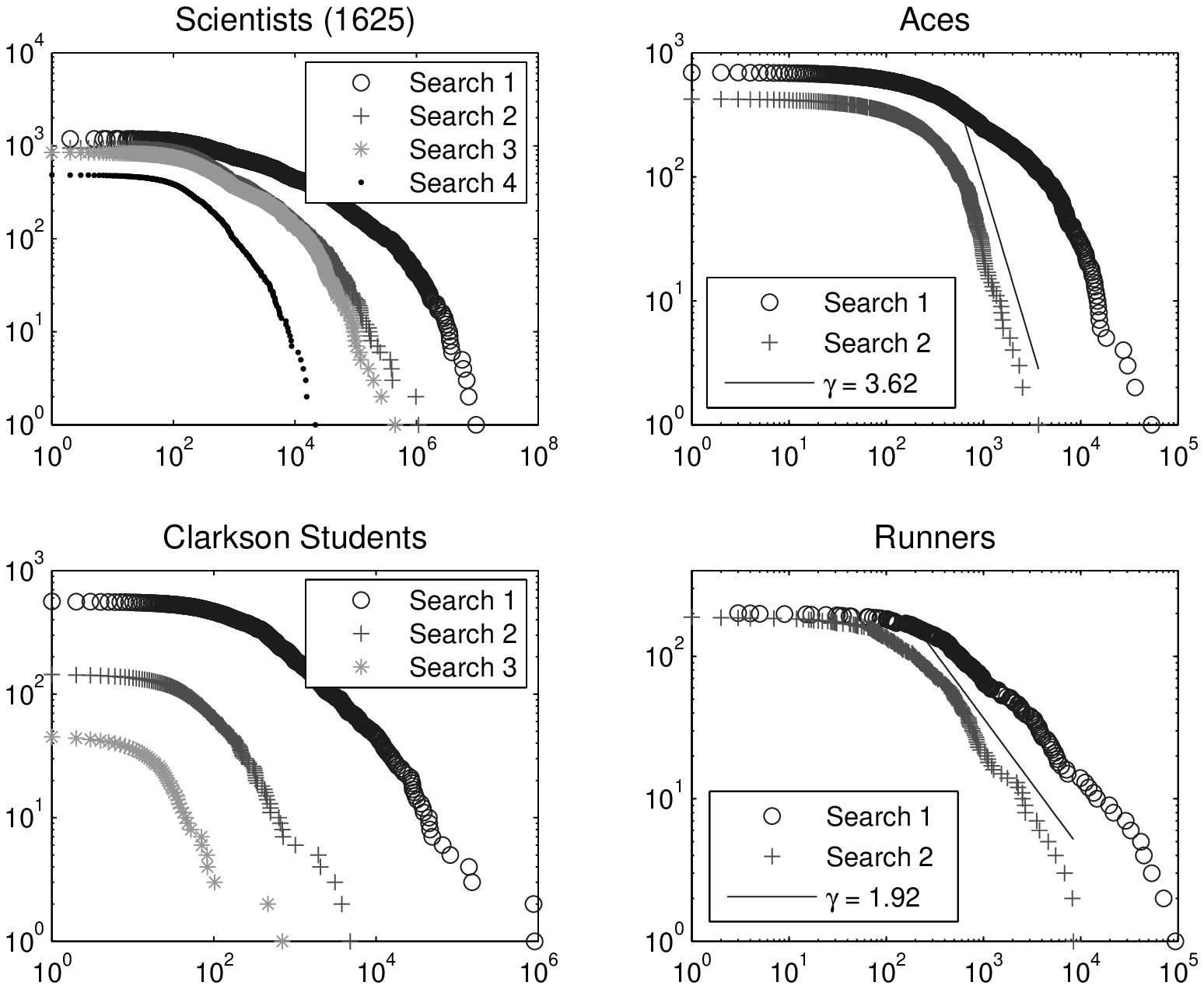}
  \caption{Rank / Frequency plots for several populations.  The horizontal axis is the number of \emph{Google} hits and the vertical axis is the rank of the (sorted) data points.  Note that straight lines (offset for clarity) will have slope $-\gamma + 1$.}
  \label{figure:ranks}
\end{figure}

The power-law exponent tends to increase as the restriction due to the lexicon increases.  This is expected because a more restrictive search will make high hit counts less frequent, increasing the slope of the tail.  Figure \ref{figure:ranks} contains \emph{rank / frequency} plots for several populations to illustrate this effect.  The plots are proportional to the empirical CDF, $P(X>x)$.  Note that individual searches which return zero hits are not shown, changing the maximum rank between lexicons.  This is most evident in the Students population: the third lexicon is very restrictive and many students garnered zero hits.

The proposed model in~\cite{originalAces} was shown to have a power-law exponent that approaches 2 asymptotically, from \emph{above}, as the number of relevant web pages citing the population in question increases over time.  The population changes in size in Table~\ref{table:results} are due to progressively restrictive lexicons and do not pertain to the same phenomenon.  On the other hand,
we are unable to account for the many instances of power-law exponents smaller than 2 observed, as any reasonable extension of the theory in~\cite{originalAces} yields powers $\gamma\geq2$.

\section{Conclusions}
A purely visual inspection of plots such as those in Figure \ref{figure:ranks} may lead one to conclude that a search is exponentially or power-law distributed, but this is misleading and subjective.  The eye will overweigh the number of data points in the tail, due to the logarithmic axes.  Objective hypothesis tests such as the KS-test must be used.

 In addition to problems with hit estimation, \texttt{OR}, etc., the choice of a lexicon has noticeable impact.  In the rank / frequency plot for the aces population in Figure \ref{figure:ranks}, the second search shows a much cleaner tail, though again this region contains less than 6\% of the aces.  All of these factors make it difficult to test theories.   For the size of populations involved, \emph{Google} hits have too much ``noise'' to accurately distinguish distributions.  



\begin{theacknowledgments}
  We are grateful to Eduard Vives for suggesting the analysis of cumulative plots by the method of maximum likelihood and to the Universidad de Granada for facilitating discussions leading to this work.  
  DbA thanks NSF grant PHY-0140094 for partial support of this work.
  \end{theacknowledgments}



{}


\end{document}